\documentclass[prl,twocolumn,final,showpacs,floatfix,superscriptaddress,nofootinbib]{revtex4}

\usepackage[utf8]{inputenc}
\usepackage{amsfonts}
\usepackage{amsmath}
\usepackage{amssymb}
\usepackage{graphicx}
\usepackage{nicefrac}
\usepackage{slashed}
\usepackage{bigstrut}
\usepackage{morefloats}
\usepackage{multirow}
\usepackage{longtable}
\usepackage{dcolumn}
\usepackage{placeins}
\usepackage{soul}
\usepackage[pageanchor]{hyperref}
\usepackage[final]{pdfpages}
\hypersetup{
        colorlinks      =true,
        urlcolor        =blue,
        linkcolor       =blue,
        citecolor       =blue,
        }
        
\allowdisplaybreaks[1]

\providecommand{\U}[1]{\protect\rule{.1in}{.1in}}

\newcommand{\up}{\uparrow}
\newcommand{\down}{\downarrow}
\DeclareMathOperator{\Tr}{Tr}

\begin{document}

\title{{\em Ab initio} lattice results for Fermi polarons in two dimensions}

\author{Shahin Bour}
\affiliation{Helmholtz-Institut f{\"u}r Strahlen- und Kernphysik (Theorie)\\
and Bethe Center for Theoretical Physics, Universit{\"a}t Bonn, 53115 Bonn,
Germany}

\author{Dean Lee}
\affiliation{Department of Physics, North Carolina State University, Raleigh,
NC 27695, USA}

\author{H.-W. Hammer}
\affiliation{Institut f{\"u}r Kernphysik, Technische Universit{\"a}t Darmstadt, D-64289 Darmstadt, Germany}
\affiliation{ExtreMe Matter Institute EMMI, GSI Helmholtzzentrum 
f{\"u}r Schwerionenforschung GmbH, D-64291 Darmstadt, Germany}

\author{Ulf-G.~Mei{\ss}ner}
\affiliation{Helmholtz-Institut f{\"u}r Strahlen- und Kernphysik (Theorie)\\
and Bethe Center for Theoretical Physics, Universit{\"a}t Bonn, 53115 Bonn,
Germany}
\affiliation{Institut f{\"u}r Kernphysik, Institute for Advanced Simulation and
J{\"u}lich Center for Hadron Physics, Forschungszentrum J{\"u}lich, D-52425
J{\"u}lich, Germany}

\begin{abstract}
We investigate the attractive Fermi polaron problem in two dimensions 
using non-perturbative Monte Carlo simulations. We introduce a new 
Monte Carlo algorithm called the impurity lattice Monte Carlo method. This
algorithm samples the path integral in a computationally efficient manner and has only small sign oscillations for 
systems with a single impurity. As a benchmark of the method, we calculate the universal polaron energy in three 
dimensions in the scale-invariant unitarity limit and find agreement with published results. We then present the first fully non-perturbative calculations of the polaron energy in two dimensions and density correlations
between the impurity and majority particles  in the limit of zero range interactions.  We find evidence for 
a smooth crossover transition from fermionic quasiparticle to molecular state as a function of interaction strength.\end{abstract}

\maketitle

One of the most interesting and fundamental problems in quantum many-body physics is 
the polaron problem, where a mobile impurity interacts with a bath of particles. With 
the advent of trapped ultracold atomic gases, the polaron problem can now be realized for both 
bosonic and fermionic baths, and also in the universal limit where the range of the 
particle interactions are negligible \cite{Bloch}. In a fermionic medium, the impurity 
can undergo a  transition and change its quantum statistics by binding fermions 
from the surrounding Fermi gas \cite{Prokof,Prokof1}. The impurity is dressed by 
fluctuations of the Fermi sea forming a quasiparticle or polaron state. But with 
increasing particle interaction strength, molecules will form by capturing one or 
even two particles from the Fermi sea, and this behavior has been shown to depend on 
the mass ratio of the two components of the Fermi gas for the 3D case \cite{Prokof,
Prokof1,Mora,Mora1,Mathy,Comb,Enss1,Nascimbene,Sommer}. In 1D, the exact analytical 
solution for equal masses shows that the polaron-molecule transition is a smooth crossover \cite{McGuire,Guan}.

In 2D the Fermi polaron 
properties have been studied using different theoretical and experimental approaches, 
and these have predicted various scenarios for the existence or absence of a 
polaron-molecule transition \cite{zolner,Vliet,Kroiss,Klawunn,Parish,Enss2,Kohl,
Zhang,Parish1}. The Fermi polaron system has been studied using diagrammatic Monte Carlo (diag MC) \cite{Vliet,Kroiss}. The diag MC method uses a worm algorithm to stochastically sample Feynman diagrams to high orders in the coupling constant. In this work we introduce 
a non-perturbative {\it ab initio} approach called impurity lattice Monte Carlo (ILMC) \cite{Elhatisari} to investigate highly-imbalanced Fermi gases. Unlike diag MC, impurity
lattice Monte Carlo directly samples the path integral and so is a fully
non-perturbative calculation.  We present calculations for the energy of
the 2D polaron and density correlations
between the impurity and majority particles as a function of interaction
strength.  Our results show evidence for a smooth
crossover from polaron to molecule.

\em Impurity lattice Monte Carlo method\em. ~The impurity lattice Monte Carlo method is a hybrid of two Monte Carlo algorithms, namely, worldline 
and auxiliary-field Monte Carlo (AFMC) algorithms. In the ILMC approach, the spacetime worldlines of each impurity are sampled explicitly while all other particles are 
handled using the auxiliary-field formalism. The impurity worldlines also function as local auxiliary fields felt by the other particles in the system. We illustrate the method by considering a $d$-dimensional many-body system 
of two-component fermions with equal masses, $m_{\up}=m_{\down}=m$ with attractive 
interactions in the zero-range limit. In our many-body system $N$ up-spin particles 
fill the Fermi sea and one down-spin particle is an impurity immersed in this Fermi 
sea. In the zero-range limit, the interaction potential can be replaced by a delta 
function interaction. Using the lattice spacing to regularize the short distance 
physics, we can write our lattice Hamiltonian as
\begin{equation}
\label{1}
H=H_{0}+C_0\sum_{\vec{n}}a^{\dagger}_{\up}(\vec{n})a^{}_{\up}
(\vec{n})a^{\dagger}_{\down}(\vec{n})a^{}_{\down}(\vec{n}),
\end{equation}
where $\vec{n}$ denote spatial lattice points on a $d$-dimensional $L^d$ periodic 
cube. The free lattice Hamiltonian $H_0$ is given by
\begin{align}
\label{2}
&H_{0}=H^{\uparrow}_{0}+H^{\downarrow}_{0}=\nonumber\\
&\tfrac{-\hbar^2}{2ma^{2}}\sum_{\mu=1}^{d}\sum_{\vec{n},i=\up,\down}
a^{\dagger}_{i}(\vec{n})\big[a^{}_{i}(\vec{n}+\hat\mu)-2a^{}_{i}(\vec{n})
+a^{}_{i}(\vec{n}-\hat\mu)\big],
\end{align}
where $a$ is the spatial lattice spacing. We fix the coupling constant $C_0$ in 
order to reproduce either the desired two-particle scattering length or the binding 
energy of a shallow dimer at infinite volume. The partition function of the system 
can be written as 
\begin{equation}
\label{3}
 \mathcal{Z} = \Tr(M^{L_t}),
\end{equation}
where $M$ is the normal-ordered transfer matrix operator
\begin{equation}
\label{4}
 M = :e^{-a_t\big[H_{0}+C_0\sum_{\vec{n}}a^{\dagger}_{\up}(\vec{n})
a^{}_{\up}(\vec{n})a^{\dagger}_{\down}(\vec{n})a^{}_{\down}(\vec{n})\big]/\hbar}:,
\end{equation}
and $a_t$ is the temporal lattice spacing.

We are interested in the system containing one down-spin particle together with $N$ up-spin particles. Let us consider one forward time step from $n_t$ 
to $n_t+1$. If the down-spin worldline remains stationary at some lattice site 
$\vec{n}$ during this time step, then there is an interaction between the impurity and the up spins. We get an effective transfer matrix for the 
up-spin particles that has the form
\begin{align}
\label{6}
M_{\uparrow}(n_t)=(1-\tfrac{da_t\hbar}{ma^{2}}):e^{-a_t\big[H^{\up}_0+\frac{C_0}
{1-\frac{da_t\hbar}{ma^{2}}}\rho_{\up}(\vec{n})\big]/\hbar}:.
\end{align}
We note the local potential generated by the down-spin impurity sitting at lattice 
site $\vec{n}$. If the down-spin worldline instead hops from one spatial lattice site 
to another, then there is no interaction between the impurity and the up spins.  Therefore the effective up-spin transfer matrix is\begin{align}
\label{7}
M_{\uparrow}(n_t)=(\tfrac{a_t\hbar}{2ma^{2}}):e^{-a_t H^{\up}_0/\hbar}:.
\end{align}
For more details on the impurity lattice Monte Carlo formalism we refer to 
Ref.~\cite{Elhatisari}.


\em Polarons in three dimensions at unitarity\em. ~As a benchmark of the ILMC method, we present simulations of polarons in the 3D unitarity limit. We define $\epsilon_{\text{p}}<0$ as the difference between ground state energy of 
the system with a single impurity compared to the system without the impurity. In the 
unitarity limit, where the $S$-wave scattering length diverges, the polaron energy 
is a universal quantity and scales with the Fermi energy, $\epsilon_{\text{p}} = 
\theta \epsilon_{\text{F}}$, where $\theta$ is a universal dimensionless number.

Using our effective up-spin transfer matrix, we compute Euclidean time projection 
amplitudes and determine the ground state energy by taking the ratio of projection amplitudes for $L_t$ and $L_t-1$ time steps in the 
limit of large $L_t$. For more details see Supplemental Material.

In order to determine the polaron energy at unitarity, we have performed simulations for several different lattice volumes $L^3$ as well as several different values for $N$, the number of up-spin particles. Since the scattering length is tuned to infinity, taking the limit of 
infinite volume at fixed particle number corresponds to taking the continuum limit with the 
interaction range going to zero.  At fixed particle number, we determine the polaron energy for 
each system at lattice volumes $6^3$, $7^3$, $8^3$, $9^3$ and $10^3$. We then apply a linear extrapolation in 
the inverse lattice spacing, the expected leading lattice-spacing dependence of corrections from the scale-invariant 
unitarity limit. See for example Ref.~\cite{Lee:2008xsa} for similar extrapolations in the unitarity 
limit. Repeating this procedure for systems with $N=15, 20, 25, 30~\text{and}~35,$ we also extrapolate 
the polaron energy to the thermodynamic limit. For the thermodynamic limit extrapolation, we perform 
a linear fit in $1/N$.

\begin{figure}[t!]
\centering
  \includegraphics[angle=-90,width=7.0cm]{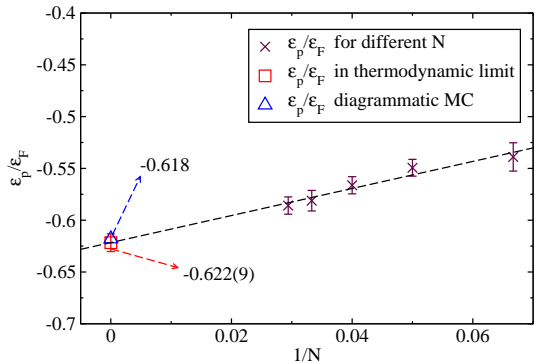}
\caption{(Color online) The 3D
unitarity limit polaron energy $\epsilon_\text{p}$ in units of the 
  Fermi energy $\epsilon_\text{F}$ as a function of the inverse
  particle number $1/N$. The square is the ILMC result
  in the thermodynamic limit, while the triangle gives the diagrammatic Monte
  Carlo result from Ref.~\cite{Prokof}.}%
\label{plot2}
\end{figure}

In Fig.~\ref{plot2} we show results for the polaron energy in the 
unitarity limit. Using the impurity lattice Monte Carlo method, we find $\theta=\epsilon
_\text{p}/\epsilon_\text{F}=-0.622(9)$. This result is in very good agreement with the 
result $\theta=-0.618$ determined in Ref.~\cite{Prokof} using diagrammatic Monte Carlo 
calculations and with the variational calculations with one and two 
particle-hole pair excitations giving $\theta = -0.6066$ \cite{Chevy} and $\theta = -0.6158$ 
\cite{Comb}, respectively. All these theoretical calculations are also consistent with the
experimental values $\theta = -0.58(5)$ \cite{Shin} and $\theta = -0.64(7)$ 
\cite{Sommer} measured in ultracold atomic gases.

\em Attractive polarons in two dimensions\em. ~We now consider attractive polarons in two 
dimensions. There is no analog of the unitarity limit in two dimensions since the only scale-invariant fixed point is in weakly interacting limit \cite{Nishida,Hammer:2009zh}.
 But there is a very interesting and important question as to whether a polaron-molecule 
transition occurs in the ground state as a function of interaction strength. At this time experiments
are not yet conclusive on the 
question of a transition \cite{Zhang,Kohl}.
The existence 
and nature of such a transition impacts the overall phase diagram for spin-imbalanced 
2D Fermi gas \cite{He,Con,Temp,Yin,parish3}. Some work using a variational approach did 
not show any ground-state transition \cite{zolner}. However later studies found evidence 
for a transition when treating molecule and polaron variational wave functions 
in a similar fashion \cite{Parish}, and similar findings have been obtained in diagrammatic Monte Carlo simulations \cite{Vliet,Kroiss}.  While these variational and diag MC studies are impressive and informative, one does not gain information about the nature of the transition itself. Separate calculations are needed to describe the fermionic polaron and the molecular state, and there is no overlapping region where both calculations are reliable. In order to remedy this situation, we use impurity lattice Monte Carlo simulations to study the non-perturbative physics of the transition in detail.

We consider again  one down-spin impurity and $N$ up-spin particles in the limit of zero-range attractive interactions. The delta function interaction is tuned according to the two-body bound state energy, $|\epsilon_B|$. Using impurity 
lattice Monte Carlo, we calculate the polaron energy as a fraction of the up-spin Fermi 
energy. We tune the coupling 
constant in order to get the two-body bound states with binding momentum $\kappa_B=\sqrt{m|
\epsilon_B|}$ equal to $0.22\hbar/a$, $0.31\hbar/a$, $0.43\hbar/a$, $0.53\hbar/a$, and 
$0.62\hbar/a$. 

We run simulations for several different lattice areas, $L^2$, and several different 
particle numbers, $N$. The lattice sizes go from $L^2 \times
L_t = 20^2 \times 100$ to $L^2 \times
L_t = 80^2 \times 700$. For each $L^2$ and $N$ we find the 
ground state energy by extrapolating to the limit $L_t\to\infty$ by fitting the Euclidean 
time projection amplitude to the asymptotic function $\epsilon_0+\alpha e^{-\delta\cdot t}$. To magnify 
the details we subtract the dimer energy in vacuum, $\epsilon_\text{B}$, from the polaron 
energies and scale by $\epsilon_\text{F}$, the majority up-spin particle Fermi energy.

\begin{figure}[t!]
\centering
  \includegraphics[width=0.35\textwidth,angle=-90]{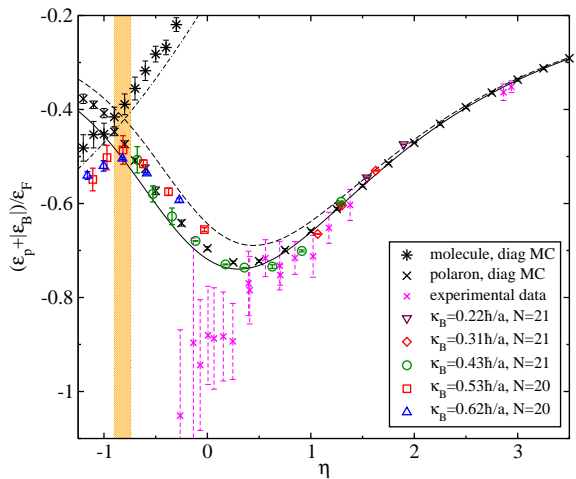}
\caption{(Color online) Ground state energy as a function of the dimensionless
parameter $\eta\equiv\frac{1}
{2}\ln\big(\frac{2\epsilon_\text{F}}{|\epsilon_\text{B}|}\big)$ in comparison
with diag MC results \cite{Vliet}. 
The experimental data is from Ref.~\cite{Kohl2}. Dashed line: polaron variational 
results  using one particle-hole (p-h) pair
\cite{Parish}. Solid line: 
polaron variational results using up to two p-h pairs
\cite{Parish1}. Dot-dashed 
line: improved molecule state variational results including one p-h pair \cite{Parish}. The vertical
band represents 
the region where the crossover transition from polaron to molecule occurs.}%
\label{plot3}
\end{figure}

In Fig.~\ref{plot3} we have plotted the subtracted-scaled polaron energy $(\epsilon_\text{p}+|\epsilon_\text{B}|)/{\epsilon_\text{F}}$ versus the dimensionless parameter $\eta\equiv\frac{1}{2}\ln
\big(\frac{2\epsilon_\text{F}}{|\epsilon_\text{B}|}\big)$, which characterizes the strength
of the interaction. The simulations are done with $N=21$ and $N=20$ up-spin particles. For comparison we have plotted the diagrammatic Monte Carlo results from Ref.~\cite{Vliet} and variational results from 
Ref.~\cite{Parish} and \cite{Parish1}. The dashed line shows polaron variational results including one particle-hole (p-h) pair
\cite{Parish}, and the solid line gives polaron variational results including up to two p-h pairs
\cite{Parish1}. The dot-dashed 
line shows the improved variational results for the molecule state including one p-h pair \cite{Parish}. We also compare with experimental data
\cite{Kohl2} presented in Ref.~\cite{Kohl} 
and find good agreement for the weak-coupling region, $\eta>1$. For the strong coupling region, $\eta<1$, the experimental
uncertainty 
increases dramatically and there seem to be non-universal systematic effects in the experimental realization that need to be corrected (see for example Ref.~\cite{Levinsen}).   

We see that for $\eta > -0.75$ the ILMC results are in excellent agreement with the fermionic polaron results obtained using diagrammatic Monte Carlo and two particle-hole variational calculations. 
For $\eta
< -0.90$, the lattice results have a similar track as the diagrammatic Monte Carlo and variational results for the molecular state.  But in this case, the lattice energies lie below the diag MC results.  This may be an indication that the non-perturbative ground wave function is a mixture of polaron and molecular components.
 In the very strong coupling limit $\eta \rightarrow -\infty$, one expects $(\epsilon_\text{p}+|\epsilon_\text{B}|)/{\epsilon_\text{F}}$
to approach $-1$ from above.
This corresponds to a tightly-bound molecule that has pulled
one up-spin from the Fermi sea and is only weakly repelled by the remaining
up spins.

Perhaps most interesting
is that the lattice results show a smooth dependence on energy in the intermediate region $-0.90<\eta<-0.75$. We interpret this as evidence for a smooth crossover from
fermionic polaron to bosonic molecule.
In the weak coupling regime, we find that the convergence to the ground state
is fastest when we use an initial state consisting of a Slater determinant
of non-interacting 
fermions. In the strong coupling regime, the convergence to the ground state
is enhanced when we use an initial state where the up-spin wave function
consists of a Gaussian wave packet for one up-spin
particle centered around the down-spin impurity, while the rest are  $N-1$ free fermions. The initial
position 
of the impurity is summed over all possible coordinates to make a translationally-invariant
state.  These two distinct optimal initial states for the weak and strong coupling limits are a clear indicator of a transition from fermionic polaron to paired molecular state as a function of coupling strength.

Despite the computational advantages of one initial state over another,
we can also obtain the same ground state energies within error bars using different initial states.
The paired Gaussian wave packet initial state does well for simulations far into the weak coupling limit when we change the size of the wave packet. At strong coupling, the optimal wave packet is fairly compact, indicating a localized pair. At weak coupling, the optimal wave packet is very large, growing as big as the lattice length $L$. 

The location of our crossover  region is in good agreement with the results of diag MC calculations 
 which have found transitions at $\eta=-0.95\pm0.15$ \cite{Vliet} and $\eta=-1.1\pm0.2$ \cite{Kroiss}. 
This result is also consistent with variational calculations \cite{Parish1} which obtains a transition in the region $-0.97<\eta<-0.80$. Our result also compares well with the experimental
result, $\eta=-0.88(20)$, obtained
after converting the experimental data
\cite{Kohl2}  performed in a quasi-2D trap to their corresponding pure 2D values \cite{Levinsen}.

We have done simulations for a wide range of particle numbers and found rapid convergence to the thermodynamic limit $N \rightarrow \infty$. In fact we find very little dependence on $N$ for $N\ge8$. The lattice results in Fig.~\ref{plot3} correspond to simulations with the largest numbers of particles, $N=21$ for weak
coupling  and $N=20$ for strong coupling. As we can see in Fig.~\ref{plot3} there is also relatively little spread in the lattice results for different values of the binding momenta. This indicates that we are close to the continuum limit of zero-range interactions, and so
corrections to the continuum limit are also numerically small.

\begin{figure}[t!]
\centering
  \includegraphics[angle=-90,width=0.46\textwidth]{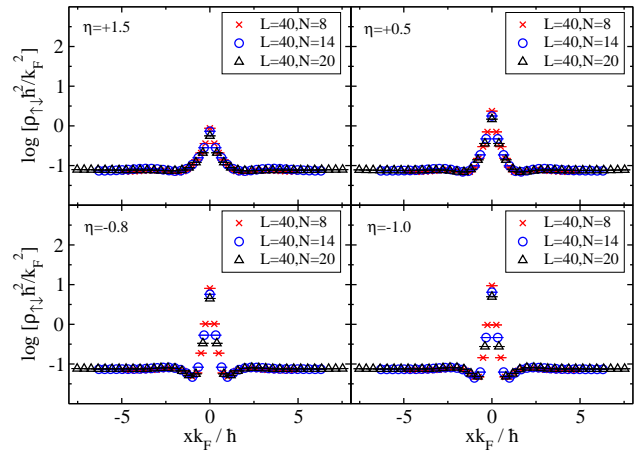}
\caption{(Color online) Plot of the density-density correlation function
$\rho_{\uparrow \downarrow}(\vec{r})$ with $\vec{r}$ measured along the $x$-axis.
We express quantities in terms of the natural length scale $\hbar
/ k_F$ and show the logarithm of the correlation function to display the
behavior at all distances.   We show results for $\eta = 1.5, 0.5, -0.8,
-1.0$ with $N = 8, 14, 20$ up-spin particles on an $L^2 = 40^2$ periodic
lattice.}
\label{density}
\end{figure}

\em Density-density correlations. \em In order to uncover the underlying nature of the polaron-molecule transition, we have used ILMC to measure the density-density correlation function between impurity and majority particles,
\begin{equation}
\rho_{\uparrow \downarrow}(\vec{r})=\int d^2r'\left< \rho_{\uparrow}(\vec{r}+\vec{r}\,')\rho_{\downarrow}(\vec{r}\,')
\right>.
\end{equation}
We have considered $N = 8, 14, 20$ up-spin particles on an $L^2 = 40^2$ periodic
lattice and interactions ranging from weak coupling to strong coupling, $\eta = 1.5, 0.5, -0.8,
-1.0$.  The results are shown in Fig.~\ref{density}. We express quantities in terms of the natural length scale $\hbar
/ k_F$ and plot the logarithm of the correlation function to show the
behavior both near and far from the central peak.

There are several interesting features apparent in Fig.~\ref{density}. First, there is little dependence on $N$, which indicates that one is already
close to the thermodynamic limit even for the small $N=8$ system. Second, although the height of the central peak depends on $\eta$, the width of the peak is approximately $\hbar
/ k_F$ and largely independent of the interaction strength. 

Third, at strong coupling we see a tall central peak and a deficit in the wings of the profile that dips below the background value.  This can be interpreted as the impurity stripping away an up-spin particle from its immediate surroundings. 
The fourth and perhaps most important finding is that there is no sign of a sharp phase transition such as a divergence of the correlation function or non-analytic dependence on the interaction strength. 
This shows that the polaron-molecule transition is a smooth crossover. 
 
\em Discussion and outlook\em.
In all of the lattice simulations presented here we have found that, with a sensible choice of initial state, the sign oscillations are mild.  This is good news for large scale ILMC simulations of other systems.  Some possible examples include alpha particles in dilute neutron gases, $\Lambda$ hyperons
in neutron and nuclear matter, and imbalanced cold atomic systems.
In these simulations one uses auxiliary-field Monte Carlo simulations for the majority particles in addition to the impurity worldline updates. If the simulations without the impurity can be done, then simulations with the impurity can also be done with little increase in sign oscillations.

At weak coupling, the impurity worldline has only a minor effect on
the eigenvalues of the $N \times N$ matrix of single-particle amplitudes, and so the determinant of the matrix remains positive.
At strong coupling we find one large positive 
eigenvalue that depends on the impurity worldline. But this large eigenvalue is positive and 
the other eigenvalues once again are largely independent of the worldline.  So the determinant remains positive in this case also. This reflects the underlying physics that at strong coupling, we have 
a tightly-bound molecule which interacts weakly with the surrounding Fermi gas. The sign problem is worst in the crossover region when we are in between 
the two extreme cases. However even in the crossover region, the sign problem remains quite manageable for the systems we have considered.

By calculating the energy of the impurity and the density-density correlations versus interaction strength, we having found evidence for a smooth crossover from polaron to molecule. 
We are now applying the ILMC method to investigate the polaron-molecule transition in three dimensions
as well as impurities in paired superfluid systems, with applications to ultracold atomic systems and alpha particles in neutron gases.


\begin{acknowledgments}
\em Acknowledgements. \em ~We gratefully acknowledge illuminating discussions with C.~Kollath, 
J.~Levinsen, M.~M.~Parish, S.~K.~Baur, J.~E.~Thomas, T.~Sch\"afer, S.~ Elhatisari, and R. Schmidt. We also 
kindly thank M.~K\"ohl, J.~Ryckebusch and ~J.~Vlietinck for sharing their data with us. Partial 
financial support was provided by the Helmholtz Association under contract HA216/EMMI, BMBF 
(grant 05P12PDFTE), and U.S. Department of Energy (DE-FG02-03ER41260). This work was further 
supported by the DFG (CRC 110),  and the Chinese Academy of Sciences (CAS)
President's International Fellowship Initiative (PIFI) (Grant No. 2015VMA076). Computational resources were provided by the J\"{u}lich Supercomputing Centre.
\end{acknowledgments}

\clearpage

\begin{figure}[!htb]
\centering
\includegraphics[scale=1]{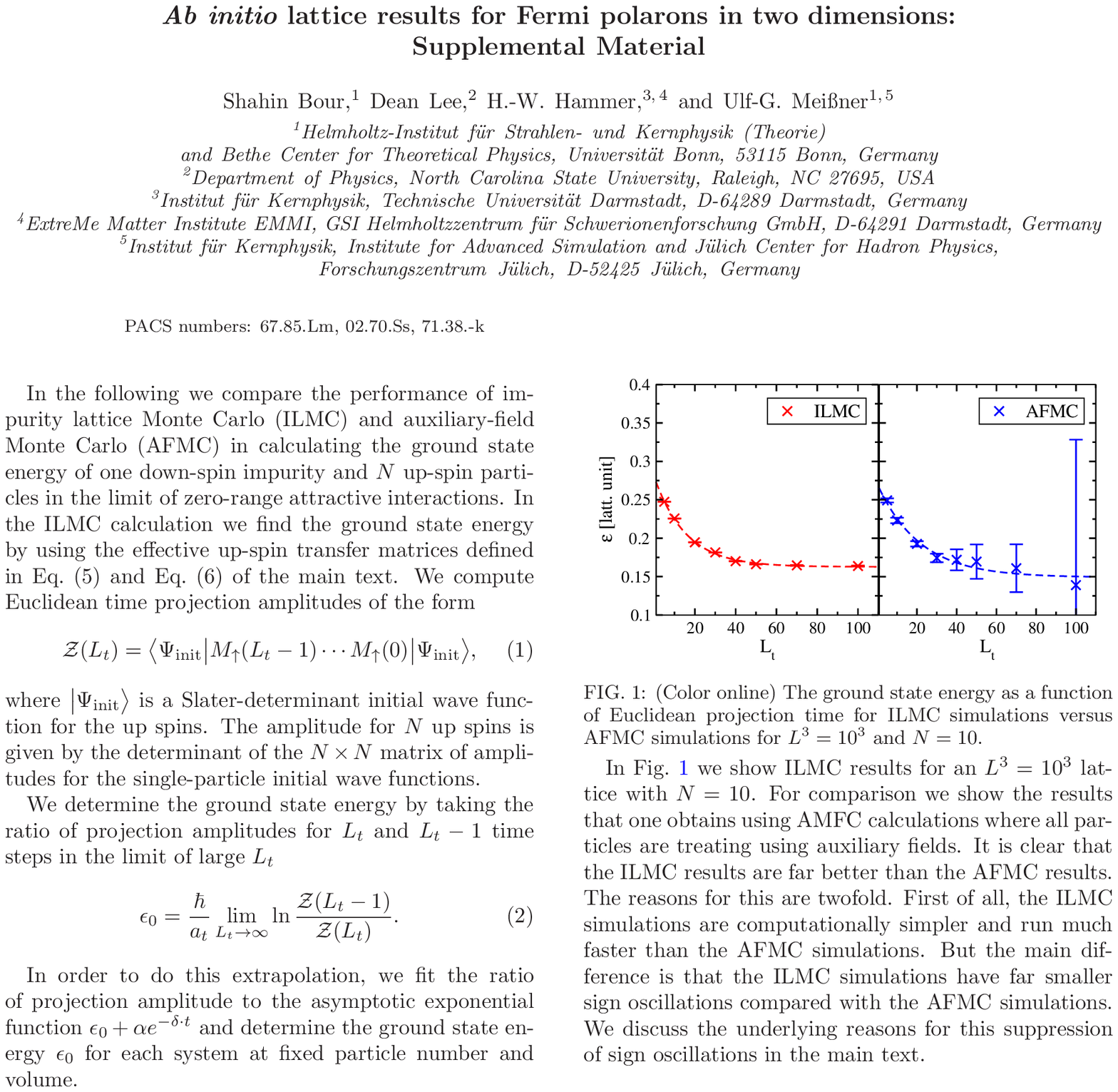}

\end{figure}

\end{document}